\documentclass[prl,aps,epsfig,twocolumn,showpacs]{revtex4}
\usepackage{epsfig,amssymb,amsmath,latexsym}
\newcommand\beq{\begin{equation}}
\newcommand\eeq{\end{equation}}
\newcommand\bea{\begin{eqnarray}}
\newcommand\eea{\end{eqnarray}}
\newcommand\bi{\begin{itemize}}
\newcommand\ei{\end{itemize}}
\newcommand\ben{\begin{enumerate}}
\newcommand\een{\end{enumerate}}
\newcommand\non{\nonumber}

\newcommand\iespace{{\textit{i.e.~}}}
\newcommand\ie{{\textit{i.e.}}}
\newcommand\odd{1{\textendash}D}
\newcommand\od{1{\textendash}D~}
\newcommand\cm{{\textsf{CM~}}}

\newcommand\cone{{${\mathbb{C}}_{1}$~}}
\newcommand\coned{{${\mathbb{C}}_{1}$}}
\newcommand\phicm{{${\mathsf{\phi_{CM}}}$~}}

\newcommand\thetacm{{${\mathsf{\theta_{CM}}}$~}}

\newcommand\ctwosq{{${\mathbb{C}}_{2}^2$~}}

\newcommand\ctwo{{${\mathbb{C}}{{_{2}}}$~}}

\newcommand\co{{${\mathbb{C}}$~}}
\newcommand\cd{{${\mathbb{C}}$}}
\newcommand\csq{{${\mathbb{C}}^2$~}}

\newcommand\cthone{{${\mathbb{C}}_{\theta _1}$~}}

\newcommand\cthtwo{{${\mathbb{C}}_{\theta _2}$~}}

\newcommand\cthreedpsq{{${\mathbb{(C}}^3_{D_P})^2$~}}

\newcommand\cthreesplus{{${\mathbb{C}}^3_{S+}$~}}

\newcommand\cthreesminus{{${\mathbb{C}}^3_{S-}$~}}
\newcommand\cthreesminusd{{${\mathbb{C}}^3_{S-}$}}
\newcommand\ro{{${\mathbb{R}}$~}}

\newcommand\rn{{${\mathbb{R}}^N$~}}

\newcommand\rs{{${\mathbb{R}}^S$~}}

\newcommand\rijd{{${\mathbb{R}}_{ij}$}}
\newcommand\cij{{${\mathbb{C}}_{ij}$~}}
\newcommand\cijd{{${\mathbb{C}}_{ij}$}}
\newcommand\dthonethtwo{{${\mathsf{D_{\theta_1 \theta_2}}}$~}}
\newcommand\dthonethtwod{{${\mathsf{D_{\theta_1 \theta_2}}}$}}
\newcommand\cik{{${\mathbb{C}}_{ik}$~}}
\newcommand\cikd{{${\mathbb{C}}_{ik}$}}
\newcommand\aone{${\mathsf{A_1~}}$}
\newcommand\aoned{${\mathsf{A_1}}$}
\newcommand\atwo{${\mathsf{A_2~}}$}
\newcommand\atwod{${\mathsf{A_2}}$}
\newcommand\athree{${\mathsf{A_3~}}$}
\newcommand\athreed{${\mathsf{A_3}}$}
\newcommand\catwo{${\mathsf{CA_2~}}$}
\newcommand\catwod{${\mathsf{CA_2}}$}

\newcommand\dnoned{${\mathsf{DN_{1}}}$}

\newcommand\dntwod{${\mathsf{DN_{2}}}$}
\newcommand\dnthree{${\mathsf{DN_{3}~}}$}
\newcommand\dnthreed{${\mathsf{DN_{3}}}$}
\newcommand\cafp{{\textsf{CA~}}}

\newcommand\nsdual{{\textsf{N$-$S~}}}
\newcommand\nsduald{{\textsf{N$-$S}}}
\newcommand\scfpd{{\textsf{S$\chi_{\pm}$}}}
\newcommand\scfp{{\textsf{S$\chi_{\pm}$~}}}
\newcommand\agfpd{{\textsf{AG}}}
\newcommand\agfp{{\textsf{AG~}}}
\newcommand\sda{{${\mathsf{SD_A}}$~}}

\newcommand\dpfp{{${\mathsf{D_P}}$~}}

\newcommand\sdpfp{{${\mathsf{SD_P}}$~}}
\newcommand\lld{{\textsf{LL}}}
\renewcommand\ll{{\textsf{LL~}}}
\newcommand\wirg{{\textsf{WIRG~}}}
\newcommand\wirgd{{\textsf{WIRG}}}
\newcommand\nsd{{\textsf{NS}}}
\newcommand\ns{{\textsf{NS~}}}
\newcommand\rgd{{\textsf{RG}}}

\newcommand\qwd{{\textsf{QW}}}
\newcommand\qw{{\textsf{QW~}}}

\newcommand\nsn{{\textsf{NSN~}}}

\newcommand\nsnd{{\textsf{NSN}}}
\newcommand\card{{\textsf{CAR}}}
\newcommand\car{{\textsf{CAR~}}}
\newcommand\ard{{\textsf{AR}}}
\newcommand\ar{{\textsf{AR~}}}
\newcommand\etal{{\hbox{{\textit\ et. al.\/}\textit\ }}}
\newcommand\etald{{\hbox{{\textit\ et. al\/}\textit\ }}}
\def\dfrac#1#2{{\displaystyle\frac{#1}{#2}}}
\newif\ifboo \boofalse

\begin{document}

\textheight=23.8cm

\title{\Large Duality between normal and superconducting junctions of multiple
quantum wires
}
\author{\bf Sourin Das$^1$ and Sumathi Rao$^2$}
\affiliation{$^1$ Centre for High Energy Physics,
 Indian Institute of Science, Bangalore 560 012, India}
\email{sourin@cts.iisc.ernet.in}
\affiliation{$^2$ \it Harish-Chandra Research Institute, Allahabad
211 019, India}
 \email{sumathi@hri.res.in}
\date{\today}
%
\begin{abstract}
We study junctions of single-channel spinless Luttinger liquids
using bosonisation. We generalize earlier studies by allowing the
junction to be superconducting and find new charge non-conserving
low energy fixed points. We establish the existence of $g
\leftrightarrow 1/g$  duality (where $g$ is the Luttinger Liquid
parameter) between the charge conserving (normal) junction and the
charge non-conserving (superconducting) junction by evaluating and
comparing the scaling dimensions of various operators around the
fixed points in both the normal and superconducting sectors of the
theory. For the most general two-wire junction, we show that there
are two conformally invariant one-parameter families of fixed points
which are also connected by a duality transformation. We also show
that the stable fixed point for the two-wire superconducting
junction corresponds to the situation where the crossed Andreev
reflection (an incoming electron is transmitted as an outgoing hole)
is perfect between the  wires. For the three-wire junction, we
study, in particular, the superconducting analogs of the chiral,
$\mathsf{D_P}$ and the disconnected fixed points obtained earlier in
the literature in the context of charge conserving three-wire
junctions.
 We show that these fixed points can be stabilized for $g < 1$
(repulsive electrons) within the superconducting sector of the
theory which makes them experimentally relevant.
\end{abstract}
%
\maketitle
\vskip .6 true cm
\section{I.\hspace{2mm} Introduction}
 \label{intro}
Recently, Y-junctions of several quasi one-dimensional (\odd)
quantum wires (\qwd) have been realized experimentally in
single-walled carbon nanotubes~\cite{fuhrer,terrones}. Junctions of
this kind are of importance for potential application in the
fabrication of quantum circuitry. Theoretically, junctions of \qw
have been studied from several points of
 view~\cite{nayak,hur,lal9,affleck1,affleck2,rao10,das2006drs,das2007drsahaprb,
 das2007drsahaepl,
chen11,egger12,pham13,safi14,moore15,yi16,kim17,furusaki18,giu20,
enss21,barnab22,barnab23,kazymyrenko24,guo25,hou} using
bosonisation, weak interaction renormalisation group (\wirgd)
methods, conformal field theory and functional renormalisation group
methods.
 The junction has also been variously taken to be enclosing a flux,
  having a resonant level,
 having a Kondo spin and having a superconductor
using one or the other techniques mentioned above.

 A comprehensive study of the junctions of three \qw enclosing
 magnetic flux was carried out by Chamon \etald~\cite{affleck1,affleck2}, where
 the wires were modeled as single channel spinless Luttinger liquids
 (\lld) and conformally invariant charge conserving boundary conditions were
identified in terms of boundary bosonic fields which had
correspondence with a host of fixed points in the theory. However,
superconducting junctions of multiple \od \qw have not been studied
in the past for the case of arbitrarily strong electron-electron
interactions.

\begin{figure*}[htb]
\begin{center}
\epsfig{figure=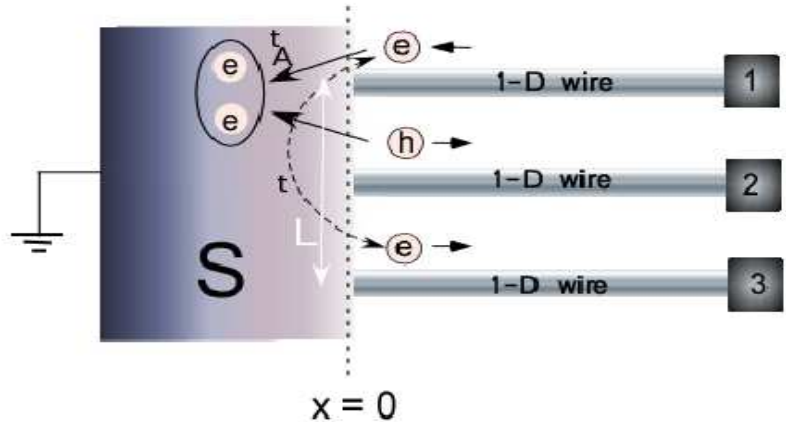,width=11cm,height=6cm}
\end{center}
\caption{Multiple wires connected to a superconducting junction
within the phase coherence length $a$ ($L < a$) of the
superconductor. The processes corresponding to an incident electron on
one wire undergoing \car (amplitude represented as $t_A$) and direct
transmission (amplitude represented as $t$) to a different wire
across the superconductor are depicted in the figure.}
\label{spfig1}
\end{figure*}

In this article, we study transport across multiple wires connected
to a superconductor as depicted in Fig.~\ref{spfig1}. In the sub-gap
region, normal reflection and transmission of the electrons cannot
occur, since charges can enter and exit the superconductor only as a
Cooper pair. But due to the proximity effect, two new processes can
occur. One is the phenomenon of Andreev reflection (\ard) in which
an electron like quasi-particle incident on
{\textsl{normal$-$superconductor}} (\nsd) junction is reflected back
as a hole along with the transfer of two electrons into the
superconductor as a Cooper pair.  The second even more interesting
process is `crossed Andreev reflection~(\card)'
~\cite{byers,feinberg, hekking1,hekking2}, whereby an electron from
one wire pairs with an electron from another wire to form a Cooper
pair and jumps into the superconductor, emitting a hole in the
second wire (note that for a singlet superconductor, the two
electrons have opposite spins). This can take place provided that
the distance between the two wires $L$ is less than or equal to the
phase coherence length of the superconductor. Thus, for an incident
electron, holes are either reflected or transmitted across the
junction, and total current conservation is taken care of by the
Cooper pairs jumping into the superconductor. However, as far as the
multiple wire system is concerned, current is not conserved. The
system is modeled as several  \od \ll connected to a superconducting
junction. We assume that the width of the superconductor between any
two wires $L \ge a$, where $a$ is the phase coherence length of the
superconductor. For simplicity, we assume that the superconductor is
a singlet. Thus spin is conserved in transport across the
superconductor and  we can confine our study to spinless \lld. For
this system, we see that the superconductor can be  modeled simply
as a (charge-violating) boundary condition on the bosons in the
wire. We also find a rich fixed point structure that generalizes the
earlier structure of fixed points  found when  multiple wires are
connected to a normal junction.

\begin{figure}[htb]
\begin{center}
\epsfig{figure=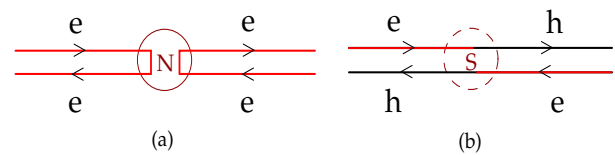,width=8.0cm,height=2cm}
\end{center}
\caption{Stable fixed points of {\textsl{(a)}} normal junction
(electron is completely reflected) and {\textsl {(b)}}
superconducting junction (electron is perfectly transmitted as a
hole).} \label{spfig2}
\end{figure}

\begin{figure}[htb]
\begin{center}
\epsfig{figure=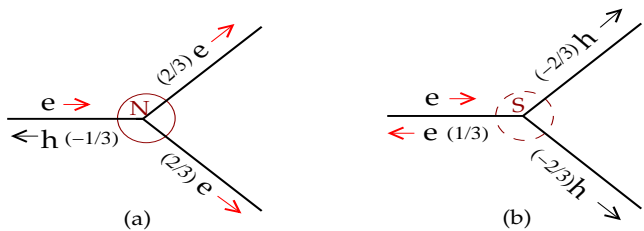,width=8.5cm,height=3cm}
\end{center}
\caption{{\textsl{(a)}} \dpfp fixed point for the normal junction.
$2/3$ charge is transmitted on each of the other wires and $-1/3$
charge is reflected, and {\textsl{(b)}} \sdpfp fixed point for the
superconducting junction. $-2/3$ charge is transmitted on each of
the other wires and $1/3$ charge is reflected.
 Note that we have
considered incoming electrons only along one of the wires.}
\label{spfig3}
\end{figure}

The superconductor explicitly violates charge conservation at the
boundary, thereby it allows for a generalization of the study of
Chamon~\etal to the charge non-conserving sector. We find that there
exists a ``normal junction$-$superconducting junction (\nsduald)"
duality given by $g \leftrightarrow 1/g$ ($g$ is the \ll parameter)
between the charge conserving (normal) and the charge non-conserving
(superconducting) sectors of the theory for junctions of any number
of \qwd.
 As a consequence of this duality, many of the fixed points that were
 unstable for the  normal junction for $g < 1$,
turn out to have stable superconducting analogs. The stability of
the fixed points mentioned here are calculated with respect to
perturbations which are within the normal sector if the fixed point
is in the normal sector and within the superconducting sector for
the fixed point in the superconducting sector. The main results
obtained in this article in the context of two-wire and three-wire
junctions are :
\ben
\item[{\textbf{(a)}}]  For the most general two-wire junction,
 we show that there are two conformally invariant
one-parameter families of fixed points which are connected to one
another via a duality transformation. We  also show that the normal
sector and the superconducting sector of the theory correspond to
two distinct points on each of two one-parameter families of
fixed points. Hence other than these special points on the two
one-parameter families, loosely speaking, the  rest of the fixed points
represent  semi-normal (semi-superconducting) junction. We find that
the stable fixed point within the superconducting sector of the
theory corresponds to a situation where an incoming electron is
completely transmitted as an outgoing hole, as shown in
Fig.~\ref{spfig2}(b). This is the crossed Andreev reflection
(\card). This fixed point is shown to be dual to the unstable
connected (perfectly transmitting) fixed point of a two-wire normal
junction due to the \nsdual duality.
%
%
\item[{\textbf{(b)}}] For the
three-wire junction, we restrict our study  to the special
cases of normal and superconducting sectors. Within each sector, the
theory of the three-wire junction effectively reduces to  the most
general theory of the two-wire junction  as in both cases it
is a theory of two independent bosonic
fields. For the three-wire junction, out  of the
three independent bosonic fields, one
 is pinned either by the charge conserving (normal) boundary condition or by
the charge non-conserving (superconducting)
 boundary condition leaving behind only two independent fields.
 Hence for the three-wire superconducting junction also,
  one gets two conformally invariant one-parameter
 families of fixed points, which are  connected to one another via a duality
 transformation.
 Of all these fixed points for the system of a
superconducting three-wire junction,
 we shall mainly focus on two which are of interest
both theoretically and experimentally :
\ben
\item[{\textbf{(i)}}] \sdpfp fixed point :
 This fixed point represents a junction with $Z_3$
symmetry between the three-wires having maximal \car between any of
the two-wires. In other words, this is a fixed point where an
incoming electron has non-zero components on all three wires as
outgoing states. $-2/3$ of the charge is transmitted on the two
other wires (hole transmission) and $1/3$ of the charge is
back-scattered (electron reflection). Note that the net change in
charge at the boundary is $e-(-2/3-2/3)e + (1/3)e = 2e$.  This  can
be identified as the charge
 non-conserving analog of the \dpfp fixed point found in
Ref.~\onlinecite{affleck2}. The
 \sdpfp fixed point is shown to be stable for $g < 1/3$
within the superconducting  sector and is identified
 as dual of the charge conserving \dpfp fixed point via the \nsdual duality.
These fixed points are shown in Fig.~\ref{spfig3}.
%
\item[{\textbf{(ii)}}] \scfp fixed point : These two fixed points,
${\mathsf{S\chi_+}}$ and ${\mathsf{S\chi_-}}$,  represent a
superconducting three-wire junction with maximally asymmetric
inter-wire \car with broken time reversal symmetry. An incoming
electron along wire 1 is transmitted as a hole in wire 2, and so on,
cyclically, as shown in Fig.~\ref{spfig3}(b) (or the other way
around).  They are   the superconducting analogs of the chiral fixed
points (Fig.~\ref{spfig3}(a))
 found earlier~\cite{lal9,affleck2,das2006drs}.
 Unlike their charge conserving analogs, these
fixed points are  stable for $1/3 < g < 1$. As can be seen from  the
stability window of \scfpd, these are  the most relevant fixed points from
the experimental point of view as they can be stabilized even for a
very weakly interacting ($g \lesssim 1$) electron gas provided the
charge conserving perturbations are weak enough.
%
\begin{figure}[htb]
\begin{center}
\epsfig{figure=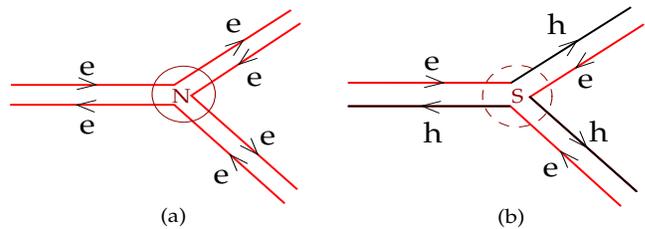,width=8.5cm,height=3cm}
\end{center}
\caption{{\textsl{(a)}} $\chi_{\pm}$ and {\textsl{(b)}} \scfp fixed
points.} \label{nsnfig4}
\end{figure}
\een
\een 


An extensive study of the renormalisation group evolution of
{several} wires connected to a superconductor was carried out very
recently in Refs.~\onlinecite{das2007drsahaprb} and
\onlinecite{das2007drsahaepl} by the present authors and A. Saha,
where conductances were studied in the Landauer-Buttiker language of
transmission and reflection of electrons. Interactions were taken
into account perturbatively using the weak interaction
renormalisation group \wirg method. But for arbitrarily  strong
inter-electron interactions, one needs to use bosonization. Also,
since the \wirg procedure is essentially a one-particle approach, it
could only access those fixed points that could be expressed
linearly in terms of fermions. To access other fixed points, one
needs to use the technique of bosonisation. Note that some of the
fixed points obtained from the fermionic \wirg method can be
identified with some of the fixed points obtained using bosonization
by taking the $g$ close to unity limit but in general this is not
true. In this paper, our aim is a comprehensive study of the system
of multiple wires connected to a superconductor, and to identify the
various fixed points of the system, only some of which were obtained
in the earlier approach.

In what follows, we first describe bosonization of superconducting
junction of $N$ number of \ll wires in Section~{II}.
In Subsections~{II A, II B and II C}, we apply this method to
single-wire, two-wire and three-wire junctions and calculate scaling
dimensions of various operators in the theory. In Subsection~{II D},
we give an expression for conductance and calculate it for various
fixed points obtained in previous subsections. Finally, we conclude
with a discussion on general issues related to the physics of \ll
junctions in Section~{III}.
\vskip .6 true cm
\section{II.\hspace{2mm} Bosonization of the superconducting
junction of \ll \qwd}
\label{boson}
The (spinless) electron field can be written in terms of bosonic
fields as,

\bea \psi (x) &=& F_O \, e^{i k_F x} \, e^{i (\,\theta(x) \,+\,
\phi(x))} ~+~ F_I e^{-i k_F x}\, e^{i(\phi(x) \,-\, \theta(x)\,)}
\nonumber \eea
\noindent where $F_O$ and $F_I$ are  Klein factors for the outgoing
and incoming  fields respectively, and $\phi(x)$ and $\theta(x)$ are
the dual bosonic fields and $k_F$ is the Fermi momentum. The wires
are modeled as spinless \ll on a half-line ($x > 0$) \ie, here we
use a folded basis for describing the junction such that all the
wires lie between $x = 0$ and $x = \infty$ and the junction is
positioned at $x=0$. Hence the action is given by

\bea {\mathsf{S}} &=& \int d\tau \int_0^\infty dx~ \sum_{i=1}^N
\Bigg[\dfrac{1}{\pi} \dot{\phi_i}\theta_i^\prime \non\\
&& + \dfrac{v}{2\pi} \left\{g \,\left(\phi_i^\prime \right)^2 +
\dfrac{1}{g}\,\left(\,\theta_i^\prime\right)^2 \right\}\Bigg] \eea
where prime (dot) stands for spatial (time) derivative and
 $ \phi_i (x,t) = (\phi_{iO} + \phi_{iI})/2$,  $\theta_i (x,t) =
(\phi_{iO} - \phi_{iI})/2$;
 ${\dot \phi_i} = (v/g)\,\theta_i^\prime$
and $ {\dot \theta_i} = (v g)\,\phi_i^\prime$.
$\phi_{iO}$ and $\phi_{iI}$ are the chiral outgoing and incoming
bosonic fields.

The action can also be written in terms of purely the $\phi_{i}$
fields or the $\theta_i$ fields and as is well-known, the two
actions are identical with the replacement of $g \leftrightarrow
1/g$. The total densities and the currents in each wire can also be
written in terms of the incoming and outgoing fields : the density
$\rho = \rho_O + \rho_I$ with $\rho_{O/I} = \pm (1/2
\pi)\,\phi_{O/I}^\prime$ and the current $j = j_O - j_I$ with
$j_{O/I} = \pm v_F(1/2\pi)\,\phi_{O/I}^\prime$. To complete the
theory, the action needs to be augmented by a boundary condition at
the origin which represents the physics at the junction.

Now following the method we used in Ref.~\onlinecite{das2006drs}, it
is possible to represent the junction in terms of a splitting matrix
\ie, we connect the incoming and the outgoing current fields as
$j_{iO} \vert_{x=0} = $ \cik $ j_{kI} \vert_{x=0}$~\footnote{The
spatial coordinate for any boundary condition is taken to be at x=0
everywhere, unless otherwise stated.} through a current splitting
matrix, \cikd. For charge conserving fixed points, the net current
flowing into the junction must be zero. Hence all charge conserving
fixed points must satisfy the constraint that $\sum_i j_i  = 0$. In
terms of the bosonic fields, this implies that $\sum_i
\phi_{iI}^\prime = -\sum_i \phi^\prime_{iO}$.
Now we can write a field splitting boundary condition as
\beq
\phi_{iO} = {\mathbb{C}}_{ij} \phi_{jI}
\eeq
 which is consistent with $j_{iO}  = $ \cik $
j_{kI} $
 and $\sum_i \phi_{iI}^\prime = - \sum_i
\phi^\prime_{iO}$. While writing the field splitting
relations in the above form, we have ignored possible integration
constants, inclusion of which makes no difference to the evaluation
of scaling dimensions of operators around various fixed points.
Hence, the current splitting and the field splitting matrices are
taken to be the same.

For the boundary condition or equivalently for the matrix \cij to
represent a fixed point, it should not flow under \rgd. This means
that it has to be scale invariant or equivalently in $1+1$ dimensions, conformally
invariant. Here, this condition simply means that the trace of the
energy-momentum tensor of the bosonic fields has to vanish at the
boundary ($x=0$). This yields the following
condition~\cite{yellowbook,polchinski1}

\beq  \sum_{i=1}^{m} {\dot \varphi_i}  ~ \varphi_i^\prime
\vert_{x=0} = 0 \eeq
\noindent where, the $\varphi_i$ are mutually independent fields
such that for a junction of $N$-wire system with $N'$ constraints,
$m = N - N'$. Hence, for two or more fields coupled to the junction,
one can have mixed boundary conditions, besides the Dirichlet
($\phi_i=0$) and
Neumann ($\phi'_i=0$)
boundary conditions. For $m = 2$  (which is the case for
the most general two-wire junction or the three-wire case in either
the purely charge conserving or the purely superconducting limit), there
are two independent families of solutions possible to the above
equation given by

 \bea { \varphi_1^\prime} = -a {\dot \varphi_2},
\quad & {\rm and} & \quad
{\varphi_2^\prime} = a {\dot \varphi_1} \\
{\dot   \varphi_1} = - a {\dot \varphi_2}, \quad & {\rm and} & \quad
{ \varphi_2^\prime} = a {\varphi_1^\prime} \eea where $a$ is a real
constant, independent of $x$ and $t$.
 In terms of the $2\times 2$ field splitting
matrices, it is easy to check that this is equivalent to taking

\beq {\mathbb{C}}_{\theta_1} =  \left (\begin{array}{cc}
~ c_1  & s_1~ \\
~  - s_1 & c_1 ~
\end{array}\right)
\quad {\rm and} \quad {\mathbb{C}}_{\theta_2} = \left
(\begin{array}{cc}
~ c_2  & s_2  ~ \\
~ s_2  & -c_2  ~
\end{array}\right)
\label{c1c2gen}
 \eeq
where $c_i= \cos \theta_i$ and  $s_i= \sin \theta_i$ are real
parameters and \cthone and \cthtwo are the field splitting matrices
at the junction for the $\varphi$ fields. The two family of
solutions are  connected via  duality transformation, which we
call  \dthonethtwod. \dthonethtwo duality can be accomplished by
either $\varphi_1 \leftrightarrow \vartheta_1$
 or $\varphi_2 \leftrightarrow \vartheta_2$. For the two-wire
 system, $\varphi_i$ and $\vartheta_i$ can be identified with the
 $\phi_i$ and $\theta_i$ respectively, $i$ being the wire index and
 \cthone and \cthtwo can be identified with the current splitting
 matrices for the two-wire system.
 For the three-wire case, $\varphi_i$ and $\vartheta_i$
 have to be taken to be linear combinations
  of $\phi_i$s and $\theta_i$s
 after imposing normal and superconducting boundary conditions.
For these one-parameter families it turns out that the incoming and
outgoing (bosonic) boundary fields satisfy the bosonic commutation
relations of the bulk given by $[\phi_{O/I}(x),\phi_{O/I}(x')] = \pm
i \pi Sgn(x-x')$, so imposing bosonic commutation relations
gives no new constraints.

 The boundary conditions may also be written in terms of the Boguliobov
 transformed free bosonic fields, $\tilde \phi_{iO/I}$ which are
defined as~\cite{das2006drs}

\bea
{\phi}_{iO} &=& \dfrac{1}{2\sqrt{g}} \left[(g+1)\,{\tilde\phi}_{iO}
-
(g-1)\,{\tilde\phi_{iI}}\right]\\
{\phi}_{iI} &=& \dfrac{1}{2\sqrt{g}} \left[-(g-1)\,{\tilde\phi}_{iO}
+ (g+1)\,{\tilde \phi}_{iI}\right]
\label{eqnseven}
\eea
For the tilde fields, the boundary condition $\phi_{iO} = $ \cij
$\phi_{jI}$ for the $N$-wire junction becomes
\beq {\tilde \phi}_{iO}  = {\mathbb{R}}_{{{ij}}}~ {\tilde \phi_{jI}}
\nonumber
 \eeq
 with
\beq
 \mathbb{R}  = \frac{[{(g+1)\,{\mathbb{C}} +
(g-1)\,\mathbb{I}}]}{[{(g-1)\,\mathbb{C} +(g+1)\,\mathbb{I}}]}
\label{rmat}
 \eeq
Thus $ \mathbb{R}$ is the matrix that connects
`free' incoming and outgoing bosonic fields whose dimensions
we know how to compute.
Now notice that when  \csq $ = \mathbb{I}$, the  above equation
simplifies to ${\tilde\phi}_{iO} =$ \cij $ {\tilde \phi}_{jI}$, but
not otherwise. This implies that, for the case of \csq $=\mathbb{I}$
both the interacting fields $(\phi_{iO/I})$ and the free fields
$(\tilde \phi_{iO/I})$ satisfy the same boundary condition. Also
note that current conservation implies that the elements of the
splitting matrix \co are real and satisfy the constraint,
%

\beq \sum_i {\mathbb{C}}_{{{ij}}} = 1 \label{normalbc} \eeq 
Furthermore the constraint that both the incoming and outgoing
fields satisfy bosonic commutation relations independently
implies~\cite{das2006drs}

\beq \sum_j {\mathbb{C}}_{ij}^2 = 1 \quad {\rm and} \quad \sum_j
{\mathbb{C}}_{ij} \mathbb{C}_{i+1,j} = 0 \label{bosonicbc} \eeq
which is essentially the same constraint that is obtained from
imposing the constraint of scale invariance, or requiring
\cij to be a fixed point.

For
the three-wire system, most of the fixed points studied in
Ref.~\onlinecite{affleck2} can be obtained as \co matrices
satisfying the above constraints. For instance, the disconnected
normal (\dnthreed) fixed point where each of the wires independently
has a Neumann boundary condition on the $\phi$ field at origin
corresponds to \co $=\mathbb{I}$ and the \dpfp fixed point has \co
matrix of the form
%

\beq {\mathbb{C}} ~=~ \left(\begin{array}{ccc}
~ -{1}/{3} & {2}/{3} & {2}/{3}~ \\
~ {2}/{3} & -{1}/{3} & {2}/{3}~\\
~ {2}/{3} & {2}/{3} & -{1}/{3}~
\end{array}\right)
\label{newfixedpoint}
 \eeq
%
It turns out that several other \cd-matrices obeying the constraints
mentioned above fall into the two one-parameter families given in
Eq.~\ref{c1c2gen} and hence can be identified as conformally invariant
fixed points. Also note that both the disconnected
\dnthree and  the above \dpfp fixed
points belong to the special class of \cij matrix for which \csq
$=\mathbb{I}$.

Physically, the disconnected \dnthree fixed point (called
${\mathsf{N}}$ in Ref.~\onlinecite{affleck2}) corresponds to a
situation where the conductance between any two-wires is zero,
whereas the \dpfp fixed point corresponds to a situation where there
is a perfect symmetry among the three-wires and the conductance
between any two-wires has the maximal value allowed by $Z_3$
symmetry. Note that this maximum is larger than the maximal
inter-wire conductance that would be allowed within a free-electron
picture for the maximally conducting $Z_3$ symmetric
case~\cite{affleck2} and this is related to the fact that for the
bosonic $Z_3$ symmetric fixed point, multi-particle scattering leads
to an enhancement of conductance as was discussed in
Ref.~\onlinecite{affleck2}.
 In Subsection~{II D}, by calculating the conductance,
 we will show that for the analogous situation in the superconducting
 sector, this is no longer true, and in fact, there is a reduction
in the conductance as compared to the free electron case. The
difference in the processes participating in the two sectors can also
be seen in Fig.~\ref{spfig3}.

 The charge conserving constraint at the junction implies that the
boundary condition on the \cm field defined as $\phi_{\mathsf{CM}} =
\sum_i \phi_i$ always has to be Neumann \ie, $\sum \phi_{iI} = \sum
\phi_{iO} \,+\,c$, where $c$ is a constant. However, in the presence
of a superconducting junction strongly coupled to the wires, there
will only  be charge non-conserving processes at the boundary (\ie,
it can either absorb or emit a Cooper-pair), and charge conserving
processes will be suppressed (at energies below the superconducting
gap). Now if we impose Dirichlet boundary condition on the \cm
field, it turns out that it gives the correct boundary condition at
the junction that converts an electron to a hole and vice-versa, and
mimics the existence of a superconductor at the junction. This leads
to new fixed points, which have not been explored in
Ref.~\onlinecite{affleck2}. This is one of the main points of our
article.

In order to establish the duality between the normal and
superconducting junctions, let us consider the case of an \ns
junction where a single \qw is connected to a superconductor (This
case was considered briefly in the appendix of
Ref.~\onlinecite{affleck2}) in the sub-gap regime.
In the limit,
when the coupling between the wire and the superconductor is strong
(\ie, there is no back-scattering of electrons), the system is in
the perfect Andreev limit and hence an incoming electron current is
completely reflected as an outgoing hole current \ie, $j_I = -j_O$.
We call this as the Andreev (\aoned) fixed point. This implies that
the boundary condition on the $\phi(x=0,t)$ field is Dirichlet (or
equivalently Neumann on the dual $\theta(x=0,t)$ field) and the
total current at the junction is given by $j = j_I - j_O = 2j_I$.
This can be easily generalized to a system of superconducting
junction of $N$-wires.

For the $N$-wire system, we must have  the sum of the incoming
electron current equal to the sum of outgoing hole current, which
means that $\sum_{i} j_{iI} + \sum_{i} j_{iO} = 0$ at the junction.
In turn, this implies that $\sum_i \rho_i = 0$, \ie, the total electron
density is zero at the junction. This is of course the correct
boundary condition as the electron density is expected to vanish at
the junction due to the finite superconducting gap. In terms of the
splitting matrix \cd, the above constraint translates into the
condition,
\beq \sum_i {\mathbb{C}}_{ij} = -1 \label{andreevbc} \eeq
in contrast to the current conserving constraint
(Eq.~\ref{normalbc})~\footnote{A similar constraint was obtained in
a different context in Ref.~\onlinecite{Bellazzini:2008mn}.}. The
other constraints coming from the bosonic commutation relations that
$\phi_{i O/I}$ have to satisfy, given by Eq.~\ref{bosonicbc}, still
remain valid. As mentioned earlier, these matrices fall into the two
one parameter families given in Eq.~\ref{c1c2gen}, thus enabling us
to identify them  as fixed points. In fact,  given a \cij matrix
representing a fixed point in the normal sector, its dual fixed
point in the superconducting sector can be obtained by transforming
\cij $\to$ $-$\cijd. It can
 be easily checked that this prescription of finding the dual fixed
 points is consistent with the constraints given by
 Eqs.~\ref{normalbc} and \ref{andreevbc}.


The duality between the charge conserving and the superconducting
boundary conditions is now obvious and can be understood physically
as follows. Current conservation implies that the net current should
be zero at the junction while, superconductivity implies that the
net electron density at the junction has to be zero, due to the existence of
the gap for single electron excitations in the superconductor. So,
in the current conserving case, the boundary condition on the \phicm
field is Neumann (or Dirichlet on \thetacm field, \ie,  $\sum_i
j_i(0,t) = 0$) while for the superconducting case, the boundary
condition is Dirichlet on \phicm field (or Neumann on \thetacm
field, \ie, $\sum_i \rho_i(0,t) = 0$). As the $\theta$ and the
$\phi$ fields have $g \leftrightarrow 1/g$ duality among themselves,
it automatically extends to the various fixed points in one sector
and their analogs in the other sector, which are obtained by
imposing further boundary conditions on the fields other than the
\cm field. We confirm this by explicitly calculating the scaling
dimension of operators corresponding to all possible perturbations
around these various fixed points.

Note however that the \nsdual duality exists over and above the
dualities that exist within each sector. For instance, within the
charge conserving sector for the two-wire system, there exists a
duality between weak back-scattering (strong tunneling) and strong
back-scattering (weak tunneling) limits with $g \leftrightarrow 1/g$
interchange. Similarly in the superconducting sector also, there
exists a duality between weak back-scattering of holes or weak
Andreev reflection (strong transmission of holes or strong \card)
and strong back-scattering of holes or strong Andreev reflection
(weak transmission of holes or weak \card) with  $g\leftrightarrow
1/g$. This essentially follows from the \dthonethtwo duality.

We will now explicitly consider the cases where there are $N=1,2$
and $3$ wires coupled to the superconductor.

\vskip 0.2cm
\subsection{A.\hspace{2mm} Single-wire
junction}\label{one}
\begin{figure}[htb]
\begin{center}
\epsfig{figure=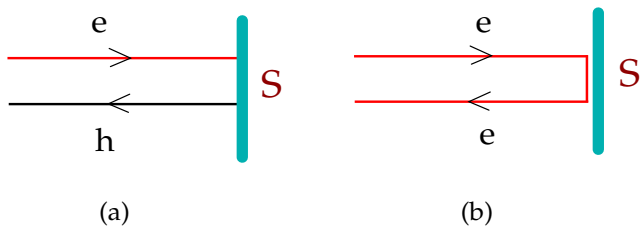,width=8.5cm,height=3.cm}
\end{center}
\caption{{\textsl{(a)}} Andreev reflection and {\textsl{(b)}} normal
reflection from a superconductor.} \label{nsnfigfive}
\end{figure}

We start with the simplest case of the \ns junction where the number
of wires, $N = 1$.  In this case, there are two single element
splitting matrices that satisfy the constraints of
Eq.~\ref{bosonicbc}, and only one of them satisfies the
superconducting constraint of \cij $= -1$. In that case the wire is
perfectly connected to the superconductor and an incoming electron
is scattered back perfectly into a hole (see Fig.~\ref{nsnfigfive}).
This is the perfect Andreev limit described before where $j_{iI} =
-j_{iO}$. The scaling dimension of the electron back-scattering
operator, $\psi_I^\dagger \psi_O$ (the subscripts $I/O$ on the
electron fields refer to incoming and outgoing branches) around this
fixed point can be easily found by bosonizing it as $\psi_I^\dagger
\psi^{}_O \sim e^{-i\phi_I} e^{i\phi_O}$. Upon writing it in terms
of the Boguliobov transformed fields, we can compute the scaling
dimension of this operator to be $2g$. Note that the back-scattering
operator we have turned on around the charge non-conserving fixed
point is charge conserving.

The other fixed point corresponds to the charge conserving case,
where the splitting matrix is \cij $=+1$. Here the incoming current
is perfectly (normal) reflected ($j_{iI}= j_{iO}$) and the wire is
completely disconnected from the superconductor. We can now turn on
a charge violating perturbation, such as the Andreev reflection
(\ard) operator, $\psi_I^{} \psi_O^{} \sim e^{i \phi_I} e^{i
\phi_O}$. The scaling dimension of this operator turns out to be
$2/g$. This establishes the  $g \leftrightarrow 1/g$ \nsdual duality
between these two cases.

\vskip .2 true cm \subsection{B.\hspace{2mm} Two-wire
junction}\label{two}
Let us now go on to case of the \nsn junction, where the number of
wires is $N = 2$. In this case, the current splitting matrix is $2
\times 2$. Unlike the previous case (\ns junction), here we find
that there are two fixed points in the superconducting sector and
they are represented by the following two matrices

\beq {\mathbb{C}}_1 =  \left (\begin{array}{cc}
~ -1 & 0~ \\
~  0 & -1~
\end{array}\right)
\quad {\rm and} \quad {\mathbb{C}}_2  = \left (\begin{array}{cc}
~ 0 & -1 ~ \\
~ -1 & 0 ~
\end{array}\right)
\label{c1c2}
 \eeq
The matrix \cone corresponds to a situation where the two-wires are
individually tuned to the  disconnected Andreev (\atwod) fixed point
(electrons are reflected back as holes) whereas the matrix \ctwo
implies perfect \car between the wires and is called the
crossed Andreev (\catwod) fixed point (electrons perfectly
transmitted as holes). As can be easily checked, \cone is a
particular case of \cthone ($\theta_1 = \pi$) and \ctwo is a
particular case of \cthtwo ($\theta_2 = -\pi/2$) given by
Eq.~\ref{c1c2gen}. It is easy to see that these two cases are
analogous to the completely reflecting (disconnected) and completely
transmitting (fully connected) cases for the normal two-wire
junction.

Let us now turn on tunneling or back-scattering operators
as perturbations around these fixed points. Around \coned, which is
fully disconnected, we switch on a \car operator which will convert
an incoming electron in one wire to an outgoing hole in another,
given by $\psi_{1I} \psi_{2O} \sim e^{i\phi_{1I}} e^{i\phi_{2O}}$.
The dimension of this operator can be computed by re-expressing the
operator in terms of the Boguliobov transformed fields.

Since the matrix \cone is just the negative of the identity matrix,
it is trivial to see that the Boguliobov transformed fields also
satisfy the same boundary conditions as the original fields. The
scaling dimension can easily be computed and it turns out to be
equal to $g$. Analogously, around the \ctwo fixed point, where  an
electron injected in the first wire gets perfectly transmitted as a
hole in the second wire, we can switch on the \ar operator,
$\psi_{iI} \psi_{iO} \sim e^{i\phi_{iI}} e^{i\phi_{iO}}$. Again the
bosonic fields can be re-expressed in terms of the Boguliobov
transformed $\tilde \phi$ fields and since \ctwosq = $\mathbb{I}$,
the tilde fields also satisfy the same boundary conditions. Here, we
find that the Andreev back-scattering operator has the dimension
$1/g$. Hence, within the superconducting sector, for repulsive
inter-electron interactions, $g < 1$, the \catwo fixed point  is a
stable fixed point (shown in Fig.~\ref{spfig2}(b)) while the fully
disconnected \atwo fixed point is unstable. This is in contrast to
the normal charge-conserving junction of two-wires, where the ``cut"
 wire (shown in Fig.~\ref{spfig2}(a))
corresponds to  stable fixed point for repulsive interactions. Again
this can also be understood in terms of the $g \leftrightarrow 1/g$
\nsdual duality.

In the above analysis, we have restricted ourselves to either
Neumann (for normal) or Dirichlet (for superconducting) boundary
conditions on the \cm field and then analysed the system, which
essentially reduces the system to a single boson ($m = 1$) problem. However,
once we allow for arbitrary charge non-conservation, then for the
two-wire system, both the \cm field $\frac{1}{{\sqrt{2}}} (\phi_1 +
\phi_2)$ and the relative field
$\frac{1}{{\sqrt{2}}}(\phi_1-\phi_2)$ enter  the picture. Hence,
the system can no longer be reduced to a single boson theory as
could be done when the  charge conserving or the superconducting
boundary condition removed the \cm field completely from the
scene.

Hence in general for a two-wire junction, we have a genuine $m = N =
2$ problem. As mentioned earlier, in terms of these two
fields, scale invariance of the boundary condition
gives us  two one-parameter family of
fixed points which are consistent with bosonic commutation
rules imposed on the incoming and outgoing fields.
The two families are connected
via duality transformation ($\phi \leftrightarrow \theta$) on either
the $\phi_1$ field or, the $\phi_2$ field, where $1$ and $2$ are
wire indices. So in conclusion, the important point to note is that
except for $\theta_1 = 0$ (``cut") and $\theta_2 = \pi/2$
(``healed") for the normal case or, $\theta_1 = \pi$ and $\theta_2 =
-\pi/2$ for the superconducting case, these fixed points belong
neither to the category of charge conserving fixed points
 nor to the category of superconducting
fixed points. A similar isolated fixed point, called
Andreev-Griffiths (\agfpd) fixed point which allowed both
superconducting and charge conserving transmissions and  reflections
was seen earlier in \wirg formalism in
Refs.~\onlinecite{das2007drsahaprb} and
\onlinecite{das2007drsahaepl} by the authors and A. Saha.

\vskip .6 true cm
\subsection{C.\hspace{2mm} Three-wire junction}\label{three}
\begin{figure}[htb]
\begin{center}
\epsfig{figure=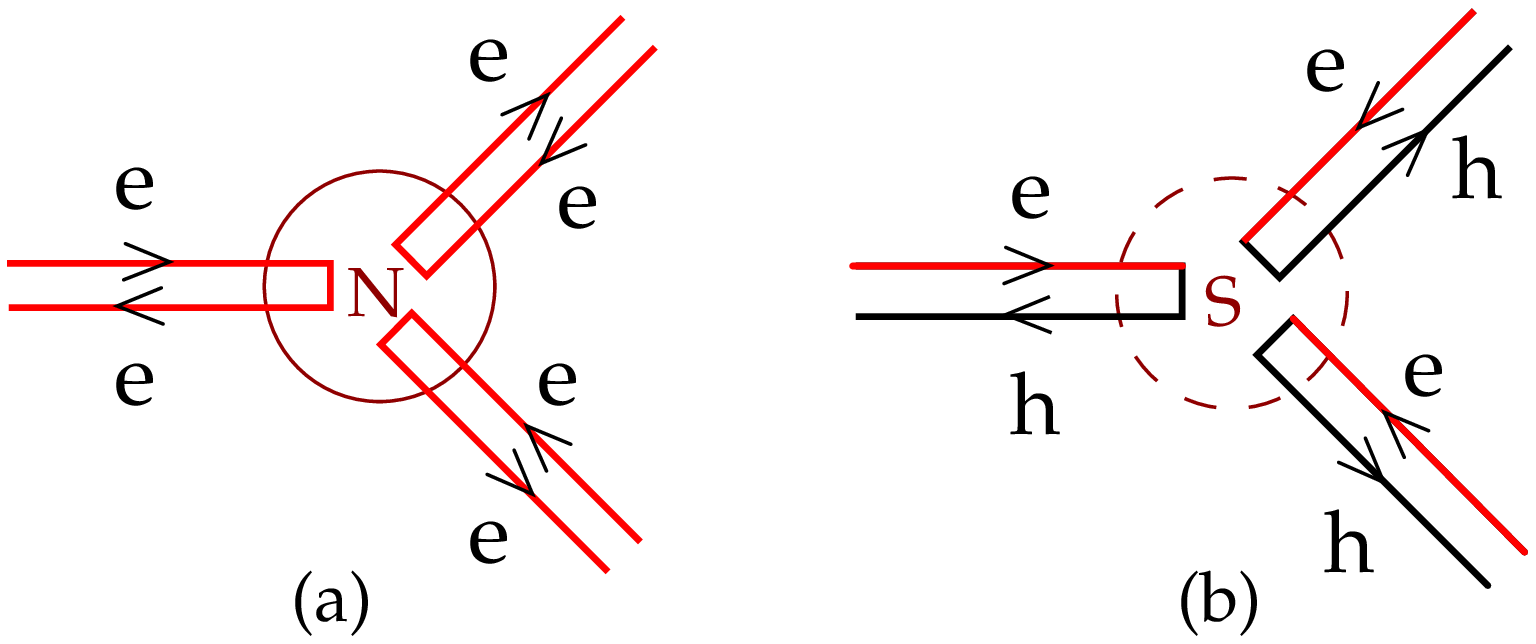,width=8.5cm,height=3.5cm}
\end{center}
\caption{ {\textsl{(a)}} \dnthree and  {\textsl{(b)}} \athree fixed
points.} \label{nsnfigsix}
\end{figure}
%
Finally, let us consider the case where there is a superconducting
junction of three \ll wires, \iespace $N = 3$ and the current
splitting matrix is $3 \times 3$. Here too, just as in the normal
three-wire case~\cite{affleck2}, we do not have a complete
classification of all the fixed points of the system in general.
However, a partial classification within the superconducting or the
normal sector can be obtained in terms of the current splitting
matrix which can be derived from the \cthone and \cthtwo matrices
given in Eq.~\ref{c1c2gen}.
For the superconducting case, it is easy to see that we will have a
fixed point corresponding to the situation
 where each individual wire is tuned to the Andreev fixed point
(see Fig.~\ref{nsnfigsix}) at the junction.
 This is the disconnected Andreev (\athreed) fixed point.
We will also have
the \sdpfp fixed point as mentioned earlier. These are the analogs
of \dnthree and \dpfp fixed points for the normal sector. We will
focus on these fixed points first. The current splitting matrices
representing the \athree and the \sdpfp are given by

\begin{widetext}
\begin{equation}
{\mathbb{C}}^3_{A_3} =  \left(\begin{array}{ccc}
~ -1 & 0 & 0~ \\
~ 0 & -1 & 0~\\
~ 0 & 0 & -1~
\end{array}\right)
\quad ~~{\textrm{and}}~~ \quad {\mathbb{C}}^3_{SD_P}
=\left(\begin{array}{ccc}
~ {1}/{3} & -{2}/{3} & -{2}/{3}~\\
~ -{2}/{3} & {1}/{3} & -{2}/{3}~\\
~ -{2}/{3} & -{2}/{3} & {1}/{3}~
\end{array}\right)
 \end{equation}
 \end{widetext}

Let us now compute the stability of these two fixed points. Around
 \athreed, the \car operator is given by
$\psi_{iI} \psi_{jO} \sim e^{i\phi_{iI}} e^{i\phi_{jO}}$ where $j
\ne i$ and $i,j=1,2,3$.  This is the same operator that we
considered in the two-wire case around the \atwo fixed point, and
the dimension of the operator of course turns out to be $g$ because
essentially it only involves tunneling between two-wires that are
disconnected from each other. This is dual to the scaling dimension
of the normal tunneling operator around the disconnected three-wire
fixed point, which is $1/g$. Thus, it gives a simple check of the
general \nsdual duality.

A more non-trivial check is  to consider the stability
of the \sdpfp fixed point. The general \nsdual duality implies that
this should be dual to the usual \dpfp fixed point of the normal
junction.
 Let us first consider the operator corresponding to \car \ie,
$\psi_{iI}^{} \psi^{}_{jO} \sim e^{i\phi_{iI}} e^{-i\phi_{jO}} $.
Since the matrix \cthreedpsq = $\mathbb{I}$, the Boguliobov
transformed bosons also satisfy the same boundary condition as the
original fields, as mentioned earlier. Hence the dimension  of the
operator can easily be computed to be $1/3g$. Now consider the
scaling dimension of the normal tunneling operator, $\psi_{iI}^{}
\psi_{jO}^\dagger \sim e^{i\phi_{iI}} e^{-i\phi_{jO}}$ around the
normal \dpfp fixed point. This has been computed in
Ref.~\onlinecite{affleck2} to be $g/3$. Thus the dimensions of these
operators are related by $g\leftrightarrow 1/g$ \nsdual duality.
Similarly, if we consider the \ar operator in each wire,
$\psi_{iI}^{} \psi^{}_{iO} \sim e^{i \phi_{iI}} e^{i \phi_{iO}}$,
its dimension can be computed to be $4/3g$. This is dual to the
dimension of the usual reflection operator, $\psi_{iI}^{}
\psi_{iO}^\dagger \sim e^{i \phi_{iI}} e^{-i \phi_{iO}}$ in a normal
junction which was earlier found to be $4 g/3$~\cite{affleck2}.

Finally, let us consider the tunneling of the incoming electron in
wire $i$ to the incoming electron in wire $j$. Here, since the
tunneling happens within the incoming channels before the electron
reaches the junction, the operator is given by $\psi_{iI}^{}
\psi_{jI}^\dagger \sim e^{i \phi_{iI}} e^{-i \phi_{jI}}$. In other
words, this is a charge conserving operator, unlike the two other
(charge violating) operators for which we calculated
 the scaling dimensions. Hence the third tunneling operator that
 we consider as a perturbation around the \sdpfp fixed point is the
same as that
 for the \dpfp fixed point. However, the scaling dimension of this
operator computed around the \sdpfp and
the \dpfp fixed points turn out to be different because the
boundary condition explicitly enters the computation of the
scaling dimensions. They turn out to be $1/g$ and $g$ respectively,
which is in
accord with the \nsdual duality.

To sum up, we find that the scaling dimensions of the three classes
of operators $(\psi_{iI}^{} \psi^{}_{jO};~ \psi^{}_{iI} \psi^{}
_{iO};~ \psi_{iI} ^{} \psi_{jI}^\dagger)$ around the \sdpfp fixed
point to be $(1/3g;~ 4/3g; ~1/g)$. These are connected by \nsdual
duality to the three classes of charge conserving operators
$(\psi_{iI}^{} \psi_{jO}^\dagger;~ \psi_{iI}^{} \psi _{iO}^\dagger;~
\psi_{iI}^{} \psi_{jI}^\dagger)$ around the \dpfp fixed point which
turn out to be  $(g/3;~ 4g/3; ~g)$. This actually exhausts the set
of all possible operators allowed by symmetry within the
superconducting sector around the \sdpfp fixed point.
 The important point to note is that
for values of g such that $g < 1/3$, all these operators are irrelevant.
So in the
 limit where the junction is tuned such that there are no
charge conserving normal
  tunnelings or reflections at the junction, this fixed
point is stable for $g < 1/3$.

 One can also perturbatively include the effect of
charge conserving tunneling and
  reflection processes  at the junction and calculate their scaling dimensions.
  Such operators are also connected by the same \nsdual duality
  transformations between the charge conserving and
  the superconducting sectors. The charge conserving tunneling
  operator,
$\psi_{iI}^{} \psi_{jO}^\dagger \sim e^{i \phi_{iI}} e^{-i
\phi_{jI}}$ between two-wires across the superconducting junction
around \sdpfp fixed point is found to have a scaling dimension $(2
g^2 + 3)/3g$. This continues to be irrelevant for $g < 1/3$; hence
it does not disturb the stability of the \sdpfp fixed point. But the
scaling dimension of the normal reflection around \sdpfp fixed point
turns out to be $2g/3$, which is relevant for $g < 1$. Hence, we
conclude that the \sdpfp fixed point is stable within the
superconducting sector for $g < 1/3$ but including the normal
reflection operator will make it flow to the disconnected charge
conserving fixed point. We can also view the \sdpfp fixed point as a
strong tunneling limit of the \car processes around the \athree
fixed point. This is analogous to the strong coupling$-$weak
coupling duality between the \dpfp and the \dnthree fixed points for
the normal three-wire system as was pointed out in
Ref.~\onlinecite{affleck2}. If we now compare the scaling dimensions
of the \car operator between the two fixed points,  these are $g$
and $1/3g$ respectively for the \athree and the \sdpfp fixed points.
In contrast, for the disconnected fixed point and \dpfp fixed point
of the normal wire, they are $1/g$ and $g/3$ respectively. Hence the
duality  $1/g \leftrightarrow g/3$ for the normal case goes over to
$g \leftrightarrow 1/3g$ in superconducting case in agreement with
the \nsdual duality.

Next we consider the \scfp fixed point. This fixed point is
described by the following two current splitting matrices given by

\begin{widetext}
\beq {\mathbb{C}}{{^3_{S+}}} = \left(\begin{array}{ccc}
~ 0  & -1 &  0~ \\
~ 0  &  0 & -1~\\
~-1  &  0 &  0 ~
\end{array}\right)
\quad~~ {\textrm{and}}~~ \quad  {\mathbb{C}}{{^3_{S-}}}  =
\left(\begin{array}{ccc}
~ 0  & 0  & -1~ \\
~-1 & 0  &  0~\\
~ 0  & -1 &  0 ~
\end{array}\right)
\label{superchiral}
 \eeq
 \end{widetext}

%
%
\begin{table*}[htb!]
\begin{center}
\begin{tabular}
{|l|l|} \hline{\sf{Normal junction}} & {\sf{Superconducting junction}}\\
\hline
(a) N=1 & (a) N=1 \\
Disconnected fixed point \dnoned, stable for $g<2$  & Andreev fixed
point \aoned, stable for $g<1/2$  \\ \hline
(b) N=2 & (b) N=2 \\
Disconnected fixed point \dntwod, stable for $g<1$   &
Andreev fixed point \atwod, stable for $g>1$  \\
Fully connected fixed point, stable for $g>1$ & Crossed Andreev
fixed point \catwod, stable for $g<1$
\\ \hline
(c) N=3 & (c) N=3 \\
Disconnected fixed point \dnthreed, stable for $g<1$ &
Andreev fixed point \athreed, stable for $g>1$ \\
Chiral fixed point, $\chi_{\pm}$, stable for $1<g<3$ &
Supercond. chiral fixed point, \scfpd, stable for $1/3<g<1$ \\
\dpfp fixed point, stable for $g>3$ & \sdpfp fixed point, stable for
$g<1/3$
\\ \hline
\end{tabular}
\caption{{Stability of some of the fixed points for different values
of the interaction parameter $g$ for $N=1,2$ and $3$ wires connected
to a normal junction and a superconducting junction are tabulated.}}
\label{table1}
\end{center}
\end{table*}
Here  the subscript $S$ stands for superconducting case and the
$+/-$ for chirality. The
 \cthreesplus fixed point corresponds to a situation where
there is perfect \car of electron from wire $1 \rightarrow 2,2
\rightarrow 3,3 \rightarrow 1$. On the other hand, \cthreesminus
corresponds to a situation where there is  perfect \car of electron
from wire $1 \rightarrow 3,3 \rightarrow 2,2 \rightarrow 1$. The
fixed point for one of these cases along with the analogous fixed
point for the normal junction is shown in Fig.~\ref{nsnfig4}. Both
these fixed points break time reversal symmetry  and depend on the
direction of the effective magnetic field through the junction.

Next we calculate the scaling dimensions of all possible operators
around these fixed points, which are the following :  (i) \ar in
each wire, $\psi_{iI}^{}  \psi_{iO}^{} \sim e^{i \phi_{iI}} e^{i
\phi_{iO}}$; (ii) \car between any two wires, $\psi^{}_{iI}
\psi^{}_{jO} \sim e^{i \phi_{iI}} e^{i \phi_{jO}}$; and (iii) normal
tunneling between the incoming chiral branches of any two wires,
$\psi_{iI}^{} \psi_{jI}^\dagger \sim e^{i \phi_{iI}} e^{-i
\phi_{jI}}$. Analogous to the normal chiral case, the scaling
dimensions of all these operators turn out to the same and are given
by $4g /(3g^2+1)$ for both \cthreesplus and \cthreesminusd. Note
that all the operators listed above are marginal for $g = 1$ and $g
= 1/3$. Now let  us compare these with the scaling dimension of all
possible operators around the chiral fixed point for the normal
junction. The current splitting matrix for the normal case just
requires the replacement of $-1$ by $1$ for both the \cthreesplus
and \cthreesminus in Eq.~\ref{superchiral}. The scaling dimensions
of all possible operators that can be switched on around either of
the fixed points are given by $4g/(3+g^2)$. It is easy to check that
the scaling dimension of the operators around the \scfp fixed point
(represented by \cthreesplus  and \cthreesminusd) is related to that
of the operators around the normal chiral fixed point by
$g\leftrightarrow 1/g$. This is as expected from the \nsdual duality
relation between the superconducting and the normal sectors.
{\textsl {However, unlike the normal chiral fixed point which is
stable for $1 < g < 3$ (attractive electrons, hence unphysical), the
\scfp fixed point is stable for $1/3 < g < 1$ (repulsive electrons);
this fact makes this fixed point experimentally relevant as this
fixed point can be stabilized even for very weakly interacting
electrons}}.

 Now
we consider the influence of charge conserving operators
corresponding to tunneling of electron ($\psi_{iI}^{}
\psi_{jO}^\dagger \sim e^{i \phi_{iI}} e^{-i \phi_{jO}}$) across the
superconducting junction between any two-wires and normal reflection
of electrons ($\psi_{iI}^{}\psi_{iO}^\dagger \sim e^{i \phi_{iI}}
e^{-i \phi_{iO}}$) within each wire. The scaling dimensions of both
these operators are given by $(2 g (1 + g^2))/(1 + 3 g^2)$ and hence
these operators are relevant for $g < 1$. This means that these
fixed points are  stable only within the superconducting sector for
$1/3 < g < 1$ but not in general.
Hence the \scfp fixed point can be relevant for experiments if one
has weakly interacting electrons but for strongly interacting
electrons ($g < 1/3$) the relevant fixed point would be the \sdpfp
fixed point as long as normal reflection at the junction is
reasonably weak.

We summarize the results of this section in Table~\ref{table1}.

\vskip .6 true cm \subsection{D.\hspace{2mm} The conductance matrix
and \nsdual duality}\label{cond:ss}

As shown explicitly in the appendix of Ref.~\onlinecite{hou}, the
conductance tensor relating the current and voltage in different
wires can be computed from the Kubo formula and can be related to
correlation functions of the  incoming and outgoing currents. These
correlation functions can be computed in terms of the  \rijd$(g)$
matrix relating the incoming and outgoing free fields. Note from
Eq.~\ref{rmat}, that this matrix is in general a function of $g$.

So far, we have discussed the dualities existing between the fixed
points in the normal and superconducting sectors in terms of scaling
dimensions of various perturbations around the fixed points. Now we
can quantify the duality in terms of the fixed point conductances of
the various fixed points in the normal sector and the corresponding
dual fixed points in the superconducting sector.

Let \rn represent the \ro matrix (which connects the `free' incoming
and outgoing fields and which has been defined in Eq.~\ref{rmat})
for a fixed point in the normal sector. Then the \ro matrix
corresponding to the dual fixed point in the superconducting sector
is given by \bea
 {\mathbb{R}}{{^S (g)}} &=& -{\mathbb{R}}{{^N(1/g)}}
 \label{rdual}
\eea  This can explicitly be checked from Eq.~\ref{rmat} when we
make the duality transformation, \cij $\to$ $-$\cijd. In terms of
the the \rn and \rs matrices, the conductance matrix is given
by~\cite{hou}
%

 \bea
{\mathbb{G}}{{^{N}}} = \dfrac{g e^2}{h} ~\left(
\mathbb{I}-{\mathbb{R}}_{}^{N}\right) ~&{\mathrm{and}}&~
{\mathbb{G}}{{^{S}}} = \dfrac{g e^2}{h}~
\left(\mathbb{I}-{\mathbb{R}}_{}^S\right)  \label{conductance} \eea
 \noindent
where $N$ and $S$ stand for normal and superconducting cases
respectively and the identity factor just represents the
incoming current along each wire.
For the two-wire  (\nsnd) junction the stable fixed
point
 within the superconducting sector for $g < 1$ was found to be the
 \atwo fixed point.
The \ro matrix is identical to the \co matrix for this case as
 \csq $=\mathbb{I}$ and hence

\beq \mathbb{G}_{A_2}^{S}  = \dfrac{g e^2}{h}~\left
(\begin{array}{cc}
~ 1 & 1 ~ \\
~ 1 & 1 ~
\end{array}\right)~.
\label{gcond}
 \eeq
This implies that if the two wires (labelled $1$ and $2$) are biased
with respect to the superconductor at voltages, $V_1$ and $V_2$ then
for an injected electron current, $(e^2/h) V_1$ the junction will
suck in an electron current equal to $(e^2/h) V_2$ from wire $2$.
Next let us consider the \sdpfp fixed point for the three-wire case.
Here also as  \csq $=\mathbb{I}$, the \ro matrix is identical to the
\co matrix and hence the conductance can be directly written down
using Eq.~\ref{conductance},

 \beq {\mathbb{G}}_{SD_{P}}^{S}  = \dfrac{ge^2}{h}~\left(\begin{array}{ccc}
~ 1 & 1 & 1~ \\
~ 1 & 1 & 1~ \\
~ 1 & 1 & 1 ~\\
\end{array}
\right)~.
\label{gsdpfp}
 \eeq
This is to be contrasted with the conductance matrix for the $Z_3$
symmetric \agfp fixed point~\cite{das2007drsahaprb} for the
three-wire system within the free-electron manifold obtained using
\wirg method given by

 \bea
{\mathbb{G}}^{S}_{\mathsf{free-electron}} &=& \dfrac{e^2}{h} ~\left(
\mathbb{I} +  \tilde {\mathbb{S}} \right)\nonumber \\
&=& \dfrac{e^2}{h} ~ \left(\begin{array}{ccc}
~ 10/9 & 4/9 & 4/9~ \\
~ 4/9 & 10/9 & 4/9~ \\
~ 4/9 & 4/9 & 10/9 ~\\
\end{array}
\right) \label{conductancewirg} \eea where, the elements of $\tilde
{\mathbb{S}}$ are obtained from the $\mathbb{S}$-matrix for the free
electron problem by taking squares of the absolute values of
individual elements themselves. This implies that $\tilde
{\mathbb{S}}$ is just the current splitting matrix for the
free electron case.

In Ref.~\onlinecite{affleck2}, it was pointed out that for the
normal junction, the diagonal conductance for the \dpfp fixed point
with Fermi liquid leads, $({\mathbb{G}}{{_{D_{P}}^{S}}})_{ii} = {4
e^2}/{3 h}$
 was larger than its free electron counterpart ${8 e^2}/{9
 h}$ for the Griffiths fixed point~\cite{lal9}.
  In contrast, we find that this is no longer true for the
  superconducting case. The free electron conductance for the
\agfp fixed point is
$10 e^2/9
  h$, which is actually larger than the diagonal conductance
for the \sdpfp fixed point with Fermi liquid leads given by
$({\mathbb{G}}{{_{SD_{P}}^{S}}})_{ii} = {e^2/h}$. So in conclusion,
electron-electron interactions lead to either
enhancement or suppression of
conductance with respect to its free electron counterpart, depending
on whether the junction is normal or superconducting.

Finally, we consider the conductance matrix for the \scfp fixed
point. For this case also, one can compute the conductance by
evaluating the $\mathbb{R}$ matrices corresponding to the
\cthreesplus and \cthreesminusd (Eq.~\ref{superchiral}) which are,
 \bea {\mathbb{R}}{{_{C^3_{S+}}}} &=& \left(
\begin{array}{ccc}
 \frac{-1+g^2}{1+3 g^2} & -\frac{2 g (1+g)}{1+3 g^2}
& -\frac{2 (-1+g) g}{1+3 g^2} \\
 -\frac{2 (-1+g) g}{1+3 g^2} & \frac{-1+g^2}{1+3 g^2}
& -\frac{2 g (1+g)}{1+3 g^2} \\
 -\frac{2 g (1+g)}{1+3 g^2} & -\frac{2 (-1+g) g}{1+3 g^2}
& \frac{-1+g^2}{1+3 g^2}
\end{array}
\right) \eea
%
%

 \bea {\mathbb{R}}_{C^3_{S-}} &=&  \left(
\begin{array}{ccc}
 \frac{-1+g^2}{1+3 g^2} & -\frac{2 (-1+g) g}{1+3 g^2}
& -\frac{2 g (1+g)}{1+3 g^2} \\
 -\frac{2 g (1+g)}{1+3 g^2} & \frac{-1+g^2}{1+3 g^2}
& -\frac{2 (-1+g) g}{1+3 g^2} \\
 -\frac{2 (-1+g) g}{1+3 g^2} & -\frac{2 g (1+g)}{1+3 g^2}
& \frac{-1+g^2}{1+3 g^2}
\end{array}
\right)\eea
%
Now the conductance matrix immediately follows from
Eq.~\ref{conductance}. Unlike the case of \sdpfp fixed point, the
$\mathbb{R}$ matrix in this case depends on the \ll parameter $g$.
Given this, one can now explicitly check the duality as defined in
Eq.~\ref{rdual}, between the chiral fixed points for the normal
junction and the \scfp fixed points in the superconducting sector, by
comparing the conductance matrix for \scfp fixed points with the
conductance matrix obtained in Ref.~\onlinecite{affleck2} for the
normal chiral fixed points.

Hence with all the explicit checks via the scaling dimensions of
operators and conductance calculations, we have established \nsdual
duality rigorously for junctions comprising
 of one, two and three-wires. In general, the \nsdual duality holds
 for junctions of any number of wires including those with $N>3$.
\vskip .6 true cm \section{III.\hspace{2mm} Discussion}
\label{conclude}

In the context of three-wires, both within the superconducting
sector and in the normal sector, there are only two independent
bosonic fields connected to the junction, since the \cm field is
constrained either by charge conservation or by the superconducting
boundary condition. As in the superconducting two-wire junction,
this implies that scale invariance of the boundary condition
 leads to two one-parameter
families of fixed points, which are related by a \dthonethtwo
duality transformation on one of the two independent fields. But for
the three wire system the \dthonethtwo duality transformation is not
a one to one map between the two one parameter families of fixed
points. For example, for the superconducting three-wire junction,
when the \dthonethtwo duality transformation is made on the
$\frac{1}{\sqrt{2}}(\phi_1 - \phi_2)$ field,  the \athree fixed
point (a representative of the \cthone family) goes to the \sda
fixed point (a $Z_3$ asymmetric fixed point
where two of the three wires are stuck at the \cafp
fixed point and the third wire is tuned to the \aone fixed point
with the junction) fixed point which is in the \cthtwo family.  But
when the duality transformation is made on the
$\frac{1}{\sqrt{6}}(\phi_1 + \phi_2 - 2\phi_3)$ field,  the
disconnected fixed point goes to the fixed point given by the matrix

\beq {\mathbb{C}}{{^3_{D_P^\prime}}}  = \left(\begin{array}{ccc}
~ -{2}/{3} & {1}/{3} & -{2}/{3}~\\
~ -{2}/{3} & -{2}/{3} & {1}/{3}~\\
~  {1}/{3} & -{2}/{3} & -{2}/{3}~
\end{array}\right)
 \eeq
 in the \cthtwo family.

 One of the most intriguing observations in the
 context of junctions of \ll in general is that in the $g \to 1$ limit, all
the fixed points in the bosonic language which have entries zero and
unity in the current splitting matrix reduce to the fermionic fixed
points which have entries zero and unity in the current splitting
matrix (obtained from the free electron ${\mathbb{S}}$-matrix) which
were obtained using \wirg formalism. The scaling dimensions of all
possible perturbations around these fixed points also  match in this
limit. But this is not generally true when the entries are not 0 and
1. For instance, the \dpfp fixed point in bosonic language is not
the same as the fermionic Griffiths fixed point in the \wirg
formalism, as neither the conductance at the fixed point, nor the
scaling dimensions of the operators around the fixed point match in
the $g \to 1$ limit even though they share the same $Z_3$
symmetry and the current splitting matrix for the \dpfp fixed point
is identical to the $\mathbb{S}$-matrix for the fermionic Griffiths
fixed point.

Also from the \nsdual duality, we should expect the analogs of the
${\mathsf{D_N}}$ and the ${\mathsf{M}}$ fixed points of
Refs.~\onlinecite{affleck1} and \onlinecite{affleck2} to exist in
the superconducting three-wire system, but a more detailed analysis
of these fixed points is beyond the scope of the present work.

 Last but
not the least it is very interesting to note that all the fixed
points that we have considered in this article are noiseless
(because there is no probabilistic partitioning of the current) even
though many of them do not correspond to the perfectly transmitting
or perfectly reflecting situations. This is unlike the fermionic
fixed point obtained using \wirg formalism, where the Griffiths
fixed point and the Andreev-Griffiths  fixed points had
probabilistic partitioning of the current and were noisy.

\vskip .6 true cm

\section*{Acknowledgements}

We thank Shamik Banerjee and Diptiman Sen for many useful
discussions. SD thanks Amit Agarwal, Leonid Glazman, Karyn Le Hur,
Volker Meden and  Alexander A. Nersesyan for valuable discussions
and Poonam Mehta for a critical reading of the manuscript. SR thanks
Dmitry Aristov and Alexander Mirlin for useful discussions. SD
acknowledges the Harish-Chandra Research Institute, India for warm
hospitality during the initial stages of this work and also
acknowledges financial support under the DST project
(SR/S2/CMP-27/2006).

\bibliographystyle{apsrev}
\bibliography{myreferences}

\begin{thebibliography}{33}
\expandafter\ifx\csname natexlab\endcsname\relax\def\natexlab#1{#1}\fi
\expandafter\ifx\csname bibnamefont\endcsname\relax
  \def\bibnamefont#1{#1}\fi
\expandafter\ifx\csname bibfnamefont\endcsname\relax
  \def\bibfnamefont#1{#1}\fi
\expandafter\ifx\csname citenamefont\endcsname\relax
  \def\citenamefont#1{#1}\fi
\expandafter\ifx\csname url\endcsname\relax
  \def\url#1{\texttt{#1}}\fi
\expandafter\ifx\csname urlprefix\endcsname\relax\def\urlprefix{URL }\fi
\providecommand{\bibinfo}[2]{#2}
\providecommand{\eprint}[2][]{\url{#2}}

\bibitem[{\citenamefont{Fuhrer et~al.}(2000)\citenamefont{Fuhrer, Nygard, Shih,
  Forero, Yoon, Mazzoni, Choi, Ihm, Louie, Zettl et~al.}}]{fuhrer}
\bibinfo{author}{\bibfnamefont{M.~S.} \bibnamefont{Fuhrer}},
  \bibinfo{author}{\bibfnamefont{J.}~\bibnamefont{Nygard}},
  \bibinfo{author}{\bibfnamefont{L.}~\bibnamefont{Shih}},
  \bibinfo{author}{\bibfnamefont{M.}~\bibnamefont{Forero}},
  \bibinfo{author}{\bibfnamefont{Y.-G.} \bibnamefont{Yoon}},
  \bibinfo{author}{\bibfnamefont{M.~S.~C.} \bibnamefont{Mazzoni}},
  \bibinfo{author}{\bibfnamefont{H.~J.} \bibnamefont{Choi}},
  \bibinfo{author}{\bibfnamefont{J.}~\bibnamefont{Ihm}},
  \bibinfo{author}{\bibfnamefont{S.~G.} \bibnamefont{Louie}},
  \bibinfo{author}{\bibfnamefont{A.}~\bibnamefont{Zettl}},
  \bibnamefont{et~al.}, \bibinfo{journal}{Science}
  \textbf{\bibinfo{volume}{288}}, \bibinfo{pages}{494} (\bibinfo{year}{2000}).

\bibitem[{\citenamefont{Terrones et~al.}(2002)\citenamefont{Terrones, Banhart,
  Grobert, Charlier, Terrones, and Ajayan}}]{terrones}
\bibinfo{author}{\bibfnamefont{M.}~\bibnamefont{Terrones}},
  \bibinfo{author}{\bibfnamefont{F.}~\bibnamefont{Banhart}},
  \bibinfo{author}{\bibfnamefont{N.}~\bibnamefont{Grobert}},
  \bibinfo{author}{\bibfnamefont{J.-C.} \bibnamefont{Charlier}},
  \bibinfo{author}{\bibfnamefont{H.}~\bibnamefont{Terrones}}, \bibnamefont{and}
  \bibinfo{author}{\bibfnamefont{P.~M.} \bibnamefont{Ajayan}},
  \bibinfo{journal}{Phys. Rev. Lett.} \textbf{\bibinfo{volume}{89}},
  \bibinfo{pages}{075505} (\bibinfo{year}{2002}).

\bibitem[{\citenamefont{Nayak et~al.}(1999)\citenamefont{Nayak, Fisher, Ludwig,
  and Lin}}]{nayak}
\bibinfo{author}{\bibfnamefont{C.}~\bibnamefont{Nayak}},
  \bibinfo{author}{\bibfnamefont{M.~P.~A.} \bibnamefont{Fisher}},
  \bibinfo{author}{\bibfnamefont{A.~W.~W.} \bibnamefont{Ludwig}},
  \bibnamefont{and} \bibinfo{author}{\bibfnamefont{H.~H.} \bibnamefont{Lin}},
  \bibinfo{journal}{Phys. Rev. B} \textbf{\bibinfo{volume}{59}},
  \bibinfo{pages}{15694} (\bibinfo{year}{1999}).

\bibitem[{\citenamefont{Hur}(2000)}]{hur}
\bibinfo{author}{\bibfnamefont{K.~L.} \bibnamefont{Hur}},
  \bibinfo{journal}{Phys. Rev. B} \textbf{\bibinfo{volume}{61}},
  \bibinfo{pages}{1853} (\bibinfo{year}{2000}).

\bibitem[{\citenamefont{Lal et~al.}(2002)\citenamefont{Lal, Rao, and
  Sen}}]{lal9}
\bibinfo{author}{\bibfnamefont{S.}~\bibnamefont{Lal}},
  \bibinfo{author}{\bibfnamefont{S.}~\bibnamefont{Rao}}, \bibnamefont{and}
  \bibinfo{author}{\bibfnamefont{D.}~\bibnamefont{Sen}},
  \bibinfo{journal}{Phys. Rev. B} \textbf{\bibinfo{volume}{66}},
  \bibinfo{pages}{165327} (\bibinfo{year}{2002}).

\bibitem[{\citenamefont{Chamon et~al.}(2003)\citenamefont{Chamon, Oshikawa, and
  Affleck}}]{affleck1}
\bibinfo{author}{\bibfnamefont{C.}~\bibnamefont{Chamon}},
  \bibinfo{author}{\bibfnamefont{M.}~\bibnamefont{Oshikawa}}, \bibnamefont{and}
  \bibinfo{author}{\bibfnamefont{I.}~\bibnamefont{Affleck}},
  \bibinfo{journal}{Phys. Rev. Lett.} \textbf{\bibinfo{volume}{91}},
  \bibinfo{pages}{206403} (\bibinfo{year}{2003}).

\bibitem[{\citenamefont{Oshikawa et~al.}(2006)\citenamefont{Oshikawa, Chamon,
  and Affleck}}]{affleck2}
\bibinfo{author}{\bibfnamefont{M.}~\bibnamefont{Oshikawa}},
  \bibinfo{author}{\bibfnamefont{C.}~\bibnamefont{Chamon}}, \bibnamefont{and}
  \bibinfo{author}{\bibfnamefont{I.}~\bibnamefont{Affleck}},
  \bibinfo{journal}{J. Stat. Mech.} \textbf{\bibinfo{volume}{0602}},
  \bibinfo{pages}{P008} (\bibinfo{year}{2006}), \eprint{cond-mat/0509675}.

\bibitem[{\citenamefont{Rao and Sen}(2004)}]{rao10}
\bibinfo{author}{\bibfnamefont{S.}~\bibnamefont{Rao}} \bibnamefont{and}
  \bibinfo{author}{\bibfnamefont{D.}~\bibnamefont{Sen}},
  \bibinfo{journal}{Phys. Rev. B} \textbf{\bibinfo{volume}{70}},
  \bibinfo{pages}{195115} (\bibinfo{year}{2004}).

\bibitem[{\citenamefont{Das et~al.}(2006)\citenamefont{Das, Rao, and
  Sen}}]{das2006drs}
\bibinfo{author}{\bibfnamefont{S.}~\bibnamefont{Das}},
  \bibinfo{author}{\bibfnamefont{S.}~\bibnamefont{Rao}}, \bibnamefont{and}
  \bibinfo{author}{\bibfnamefont{D.}~\bibnamefont{Sen}},
  \bibinfo{journal}{Phys. Rev. B} \textbf{\bibinfo{volume}{74}},
  \bibinfo{eid}{045322} (\bibinfo{year}{2006}).

\bibitem[{\citenamefont{Das et~al.}(2008{\natexlab{a}})\citenamefont{Das, Rao,
  and Saha}}]{das2007drsahaprb}
\bibinfo{author}{\bibfnamefont{S.}~\bibnamefont{Das}},
  \bibinfo{author}{\bibfnamefont{S.}~\bibnamefont{Rao}}, \bibnamefont{and}
  \bibinfo{author}{\bibfnamefont{A.}~\bibnamefont{Saha}},
  \bibinfo{journal}{Phys. Rev. B} \textbf{\bibinfo{volume}{77}},
  \bibinfo{eid}{155418} (\bibinfo{year}{2008}{\natexlab{a}}).

\bibitem[{\citenamefont{Das et~al.}(2008{\natexlab{b}})\citenamefont{Das, Rao,
  and Saha}}]{das2007drsahaepl}
\bibinfo{author}{\bibfnamefont{S.}~\bibnamefont{Das}},
  \bibinfo{author}{\bibfnamefont{S.}~\bibnamefont{Rao}}, \bibnamefont{and}
  \bibinfo{author}{\bibfnamefont{A.}~\bibnamefont{Saha}},
  \bibinfo{journal}{Europhys. Lett.} \textbf{\bibinfo{volume}{81}},
  \bibinfo{pages}{67001} (\bibinfo{year}{2008}{\natexlab{b}}).

\bibitem[{\citenamefont{Chen et~al.}(2002)\citenamefont{Chen, Trauzettel, and
  Egger}}]{chen11}
\bibinfo{author}{\bibfnamefont{S.}~\bibnamefont{Chen}},
  \bibinfo{author}{\bibfnamefont{B.}~\bibnamefont{Trauzettel}},
  \bibnamefont{and} \bibinfo{author}{\bibfnamefont{R.}~\bibnamefont{Egger}},
  \bibinfo{journal}{Phys. Rev. Lett.} \textbf{\bibinfo{volume}{89}},
  \bibinfo{pages}{226404} (\bibinfo{year}{2002}).

\bibitem[{\citenamefont{Egger et~al.}(2003)\citenamefont{Egger, Trauzettel,
  Chen, and Siano}}]{egger12}
\bibinfo{author}{\bibfnamefont{R.}~\bibnamefont{Egger}},
  \bibinfo{author}{\bibfnamefont{B.}~\bibnamefont{Trauzettel}},
  \bibinfo{author}{\bibfnamefont{S.}~\bibnamefont{Chen}}, \bibnamefont{and}
  \bibinfo{author}{\bibfnamefont{F.}~\bibnamefont{Siano}},
  \bibinfo{journal}{New Journal of Physics} \textbf{\bibinfo{volume}{5}},
  \bibinfo{pages}{117.1} (\bibinfo{year}{2003}).

\bibitem[{\citenamefont{Pham et~al.}(2003)\citenamefont{Pham, Pi\'echon, Imura,
  and Lederer}}]{pham13}
\bibinfo{author}{\bibfnamefont{K.-V.} \bibnamefont{Pham}},
  \bibinfo{author}{\bibfnamefont{F.}~\bibnamefont{Pi\'echon}},
  \bibinfo{author}{\bibfnamefont{K.-I.} \bibnamefont{Imura}}, \bibnamefont{and}
  \bibinfo{author}{\bibfnamefont{P.}~\bibnamefont{Lederer}},
  \bibinfo{journal}{Phys. Rev. B} \textbf{\bibinfo{volume}{68}},
  \bibinfo{pages}{205110} (\bibinfo{year}{2003}).

\bibitem[{\citenamefont{Safi et~al.}(2001)\citenamefont{Safi, Devillard, and
  Martin}}]{safi14}
\bibinfo{author}{\bibfnamefont{I.}~\bibnamefont{Safi}},
  \bibinfo{author}{\bibfnamefont{P.}~\bibnamefont{Devillard}},
  \bibnamefont{and} \bibinfo{author}{\bibfnamefont{T.}~\bibnamefont{Martin}},
  \bibinfo{journal}{Phys. Rev. Lett.} \textbf{\bibinfo{volume}{86}},
  \bibinfo{pages}{4628} (\bibinfo{year}{2001}).

\bibitem[{\citenamefont{Moore and Wen}(2002)}]{moore15}
\bibinfo{author}{\bibfnamefont{J.~E.} \bibnamefont{Moore}} \bibnamefont{and}
  \bibinfo{author}{\bibfnamefont{X.-G.} \bibnamefont{Wen}},
  \bibinfo{journal}{Phys. Rev. B} \textbf{\bibinfo{volume}{66}},
  \bibinfo{pages}{115305} (\bibinfo{year}{2002}).

\bibitem[{\citenamefont{Yi}(2002)}]{yi16}
\bibinfo{author}{\bibfnamefont{H.}~\bibnamefont{Yi}}, \bibinfo{journal}{Phys.
  Rev. B} \textbf{\bibinfo{volume}{65}}, \bibinfo{pages}{195101}
  (\bibinfo{year}{2002}).

\bibitem[{\citenamefont{Kim et~al.}(2004)\citenamefont{Kim, Vishveshwara, and
  Fradkin}}]{kim17}
\bibinfo{author}{\bibfnamefont{E.-A.} \bibnamefont{Kim}},
  \bibinfo{author}{\bibfnamefont{S.}~\bibnamefont{Vishveshwara}},
  \bibnamefont{and} \bibinfo{author}{\bibfnamefont{E.}~\bibnamefont{Fradkin}},
  \bibinfo{journal}{Phys. Rev. Lett.} \textbf{\bibinfo{volume}{93}},
  \bibinfo{pages}{266803} (\bibinfo{year}{2004}).

\bibitem[{\citenamefont{Furusaki}(2005)}]{furusaki18}
\bibinfo{author}{\bibfnamefont{A.}~\bibnamefont{Furusaki}},
  \bibinfo{journal}{J. Phys. Soc. Japan} \textbf{\bibinfo{volume}{74}},
  \bibinfo{pages}{73} (\bibinfo{year}{2005}).

\bibitem[{\citenamefont{Giuliano and Sodano}(2005)}]{giu20}
\bibinfo{author}{\bibfnamefont{D.}~\bibnamefont{Giuliano}} \bibnamefont{and}
  \bibinfo{author}{\bibfnamefont{P.}~\bibnamefont{Sodano}},
  \bibinfo{journal}{Nucl. Phys. B} \textbf{\bibinfo{volume}{711}},
  \bibinfo{pages}{480} (\bibinfo{year}{2005}).

\bibitem[{\citenamefont{Enss et~al.}(2005)\citenamefont{Enss, Meden,
  Andergassen, Barnab\'{e}-Th\'{e}riault, Metzner, and
  Sch\"{o}nhammer}}]{enss21}
\bibinfo{author}{\bibfnamefont{T.}~\bibnamefont{Enss}},
  \bibinfo{author}{\bibfnamefont{V.}~\bibnamefont{Meden}},
  \bibinfo{author}{\bibfnamefont{S.}~\bibnamefont{Andergassen}},
  \bibinfo{author}{\bibfnamefont{X.}~\bibnamefont{Barnab\'{e}-Th\'{e}riault}},
  \bibinfo{author}{\bibfnamefont{W.}~\bibnamefont{Metzner}}, \bibnamefont{and}
  \bibinfo{author}{\bibfnamefont{K.}~\bibnamefont{Sch\"{o}nhammer}},
  \bibinfo{journal}{Phys. Rev. B} \textbf{\bibinfo{volume}{71}},
  \bibinfo{eid}{155401} (\bibinfo{year}{2005}).

\bibitem[{\citenamefont{Barnab\'{e}-Th\'{e}riault
  et~al.}(2005{\natexlab{a}})\citenamefont{Barnab\'{e}-Th\'{e}riault, Sedeki,
  Meden, and Sch\"{o}nhammer}}]{barnab22}
\bibinfo{author}{\bibfnamefont{X.}~\bibnamefont{Barnab\'{e}-Th\'{e}riault}},
  \bibinfo{author}{\bibfnamefont{A.}~\bibnamefont{Sedeki}},
  \bibinfo{author}{\bibfnamefont{V.}~\bibnamefont{Meden}}, \bibnamefont{and}
  \bibinfo{author}{\bibfnamefont{K.}~\bibnamefont{Sch\"{o}nhammer}},
  \bibinfo{journal}{Phys. Rev. Lett.} \textbf{\bibinfo{volume}{94}},
  \bibinfo{eid}{136405} (\bibinfo{year}{2005}{\natexlab{a}}).

\bibitem[{\citenamefont{Barnab\'{e}-Th\'{e}riault
  et~al.}(2005{\natexlab{b}})\citenamefont{Barnab\'{e}-Th\'{e}riault, Sedeki,
  Meden, and Sch\"{o}nhammer}}]{barnab23}
\bibinfo{author}{\bibfnamefont{X.}~\bibnamefont{Barnab\'{e}-Th\'{e}riault}},
  \bibinfo{author}{\bibfnamefont{A.}~\bibnamefont{Sedeki}},
  \bibinfo{author}{\bibfnamefont{V.}~\bibnamefont{Meden}}, \bibnamefont{and}
  \bibinfo{author}{\bibfnamefont{K.}~\bibnamefont{Sch\"{o}nhammer}},
  \bibinfo{journal}{Phys. Rev. B} \textbf{\bibinfo{volume}{71}},
  \bibinfo{eid}{205327} (\bibinfo{year}{2005}{\natexlab{b}}).

\bibitem[{\citenamefont{Kazymyrenko and Dou\c{c}ot}(2005)}]{kazymyrenko24}
\bibinfo{author}{\bibfnamefont{K.}~\bibnamefont{Kazymyrenko}} \bibnamefont{and}
  \bibinfo{author}{\bibfnamefont{B.}~\bibnamefont{Dou\c{c}ot}},
  \bibinfo{journal}{Phys. Rev. B} \textbf{\bibinfo{volume}{71}},
  \bibinfo{eid}{075110} (\bibinfo{year}{2005}).

\bibitem[{\citenamefont{Guo and White}(2006)}]{guo25}
\bibinfo{author}{\bibfnamefont{H.}~\bibnamefont{Guo}} \bibnamefont{and}
  \bibinfo{author}{\bibfnamefont{S.~R.} \bibnamefont{White}},
  \bibinfo{journal}{Phys. Rev. B} \textbf{\bibinfo{volume}{74}},
  \bibinfo{eid}{060401} (\bibinfo{year}{2006}).

\bibitem[{\citenamefont{Hou and Chamon}(2008)}]{hou}
\bibinfo{author}{\bibfnamefont{C.-Y.} \bibnamefont{Hou}} \bibnamefont{and}
  \bibinfo{author}{\bibfnamefont{C.}~\bibnamefont{Chamon}},
  \bibinfo{journal}{Phys. Rev. B} \textbf{\bibinfo{volume}{77}},
  \bibinfo{pages}{155422} (\bibinfo{year}{2008}).

\bibitem[{\citenamefont{Byers and Flatt\'e}(1995)}]{byers}
\bibinfo{author}{\bibfnamefont{J.~M.} \bibnamefont{Byers}} \bibnamefont{and}
  \bibinfo{author}{\bibfnamefont{M.~E.} \bibnamefont{Flatt\'e}},
  \bibinfo{journal}{Phys. Rev. Lett.} \textbf{\bibinfo{volume}{74}},
  \bibinfo{pages}{306} (\bibinfo{year}{1995}).

\bibitem[{\citenamefont{Deutscher and Feinberg}(2000)}]{feinberg}
\bibinfo{author}{\bibfnamefont{G.}~\bibnamefont{Deutscher}} \bibnamefont{and}
  \bibinfo{author}{\bibfnamefont{D.}~\bibnamefont{Feinberg}},
  \bibinfo{journal}{App. Phys. Lett.} \textbf{\bibinfo{volume}{76}},
  \bibinfo{pages}{487} (\bibinfo{year}{2000}).

\bibitem[{\citenamefont{{Falci} et~al.}(2001)\citenamefont{{Falci}, {Feinberg},
  and {Hekking}}}]{hekking1}
\bibinfo{author}{\bibfnamefont{G.}~\bibnamefont{{Falci}}},
  \bibinfo{author}{\bibfnamefont{D.}~\bibnamefont{{Feinberg}}},
  \bibnamefont{and} \bibinfo{author}{\bibfnamefont{F.~W.~J.}
  \bibnamefont{{Hekking}}}, \bibinfo{journal}{Europhys. Lett.}
  \textbf{\bibinfo{volume}{54}}, \bibinfo{pages}{255} (\bibinfo{year}{2001}).

\bibitem[{\citenamefont{{Bignon} et~al.}(2004)\citenamefont{{Bignon}, {Houzet},
  {Pistolesi}, and {Hekking}}}]{hekking2}
\bibinfo{author}{\bibfnamefont{G.}~\bibnamefont{{Bignon}}},
  \bibinfo{author}{\bibfnamefont{M.}~\bibnamefont{{Houzet}}},
  \bibinfo{author}{\bibfnamefont{F.}~\bibnamefont{{Pistolesi}}},
  \bibnamefont{and} \bibinfo{author}{\bibfnamefont{F.~W.~J.}
  \bibnamefont{{Hekking}}}, \bibinfo{journal}{Europhys. Lett.}
  \textbf{\bibinfo{volume}{67}}, \bibinfo{pages}{110} (\bibinfo{year}{2004}).

\bibitem[{\citenamefont{Francesco et~al.}(1997)\citenamefont{Francesco,
  Mathieu, and Senechal}}]{yellowbook}
\bibinfo{author}{\bibfnamefont{P.~D.} \bibnamefont{Francesco}},
  \bibinfo{author}{\bibfnamefont{P.}~\bibnamefont{Mathieu}}, \bibnamefont{and}
  \bibinfo{author}{\bibfnamefont{D.}~\bibnamefont{Senechal}},
  \emph{\bibinfo{title}{Conformal Field Theory,}}
  (\bibinfo{publisher}{Springer-Verlag, New York, USA}, \bibinfo{year}{1997}).

\bibitem[{\citenamefont{Polchinski}(1998)}]{polchinski1}
\bibinfo{author}{\bibfnamefont{J.}~\bibnamefont{Polchinski}},
  \emph{\bibinfo{title}{String Theory, Vol. 1 : An Introduction to the Bosonic
  String,}} (\bibinfo{publisher}{Cambridge University Press, Cambridge, UK},
  \bibinfo{year}{1998}).

\bibitem[{\citenamefont{Bellazzini et~al.}(2008)\citenamefont{Bellazzini,
  Burrello, Mintchev, and Sorba}}]{Bellazzini:2008mn}
\bibinfo{author}{\bibfnamefont{B.}~\bibnamefont{Bellazzini}},
  \bibinfo{author}{\bibfnamefont{M.}~\bibnamefont{Burrello}},
  \bibinfo{author}{\bibfnamefont{M.}~\bibnamefont{Mintchev}}, \bibnamefont{and}
  \bibinfo{author}{\bibfnamefont{P.}~\bibnamefont{Sorba}}
  (\bibinfo{year}{2008}), \eprint{arXiv:0801.2852 [hep-th]}.

\end{thebibliography}

\end{document}